\documentclass[11pt]{amsart}
mm \hoffset -16 true mm

\begin{document}
\title{Local channels preserving maximal entanglement or Schmidt number}
\author{Yu Guo}
 \address[Yu Guo]{Department of
Mathematics, Shanxi Datong University, Datong 037009, China;}
\address{Institute of Optoelectronic Engineering, Taiyuan
University of Technology, Taiyuan 030024, China.}\email[Yu
Guo]{guoyu3@yahoo.com.cn}

\author{Zhaofang Bai}
\address[Zhaofang Bai]{School of Mathematical Sciences, Xiamen University, Xiamen,
361005, P. R. China.} \email[Zhaofang Bai(Corresponding
author)]{zhaofangbai@yahoo.com.cn}
\author{Shuanping Du}
\address[Shuanping Du]{School of Mathematical Sciences, Xiamen University, Xiamen, 361005,
 P. R. China.} \email[Shuanping
Du]{shuanpingdu@yahoo.com}
\thanks{This work is partially supported by Natural Science Foundation of
China (Grants No.11171249, Grants No.11001230, Grants
No.11101250), China Postdoctoral Science Foundation Funded Project (2012M520603)
and Research start-up fund for Doctors of Shanxi Datong University
(No.2011-B-01).}
\thanks{This paper is in final form and no version of it will be submitted for
publication elsewhere.}

\begin{abstract}
Maximal entanglement and Schmidt number  play an important role in
various quantum information tasks. In this paper, it is shown that
a local channel preserves maximal entanglement state(MES) or
preserves pure states with Schmidt number $r$($r$ is a fixed
integer) if and only if it is a local unitary operation.
\end{abstract}

\maketitle 

\section{Introduction}
Quantum correlations, including the entanglement and the quantum
discord, are useful resources that play a fundamental role in
various quantum informational processes. As is well-known, the
entanglement is non-increasing under local operations and
classical communications (LOCC). Especially the entanglement
cannot be created from a separable state using only LOCCs. This
property of entanglement is characteristic for its various
quantitative measures. However, quantum correlation can be created
by local operation from some initially classical states. Recently,
some progress has been made on attacking this issue. In
Refs.\cite{Hu,Yusixia}, two group of authors proved that the
necessary and sufficient condition for a local channel to create
quantum correlation is not a commutativity-preserving channel.
Mathematically, the authors of \cite{Hu,Yusixia} were to study
what kind of local channels preserve classical states which are
usually viewed as the states  with the least quantum correlation.
It is natural to ask when the local channels preserve states with
the maximum quantum correlation. Maximal entanglement (ME) can be
viewed as the most strong quantum correlation. It is especially
important both experimentally and theoretically
\cite{Horodecki1,Guhne,Feishaoming,Nielsen}. Our  first aim in
this note is to give the necessary and sufficient condition for a
local channel preserving maximal entanglement.

Another line, the initiation of this note is also inspired by the
linear preserve problem. The study of linear preserver problems
has a long history and  growing research interest. It concerns the
characterization of maps on matrices or operators with special
properties. For example, Frobenius \cite{Fro} showed that a linear
operator $\Phi: \mathcal {M}_n\rightarrow \mathcal {M}_n$
satisfies $det(\Phi(A)) = det(A)$ for all $A \in \mathcal{M}_n$ if
and only if there are $M,N \in \mathcal{ M}_n$ with $det(MN) = 1$
such that $\Phi$ has the form
\begin{eqnarray} A \rightarrow MAN
\quad \text{or} \quad  A \rightarrow MA^tN,\end{eqnarray} where
$\mathcal{M}_n$ denotes the set of $n \times n$ complex matrices.
Clearly, a map of the form (1) is linear and leaves the
determinant function invariant. It is interesting that a linear
map preserving the determinant function must be of this form. One
may see  \cite{LiCK, Mol}and its references for results on linear
preserver problems.

In quantum information theory, it is well known that a quantum
channel on the system is described by a trace-preserving
completely positive linear map $\Lambda:~
\mathcal{T}(H)\rightarrow \mathcal{T}(H)$ (${\mathcal T}(H)$
denotes the set of  trace class operators acting on a Hilbert
space $H$) that admits a form of Kraus operator representation,
i.e.,
\begin{eqnarray}
\Lambda(\cdot)=\sum\limits_{i}X_i(\cdot)X_i^\dag
\end{eqnarray}
where, $X_i\in\mathcal{B}(H)$ (the set of all bounded linear
operators on $H$), $\sum\limits_{i}X_i^\dag X_i=I$ (the
representation is not unique), and the series converges in the
trace norm topology in the case of infinite sum. A channel
$\Lambda$ on the bipartite system A+B is called a local channel if
$\Lambda=\Lambda_a\otimes\Lambda_b$, where $\Lambda_{a/b}$ is
quantum channel on subsystem A/B. Our problem can be rewritten as
to study such maps (local channels) preserving maximum entangle
state(MES). It is obvious  that $\rho$ is a MES implies
$U_A\otimes U_B\rho U_A^\dag\otimes U_B^\dag$ is maximally
entangled, where $U_A$ and $U_B$ are unitary operators acting on
$H_A$ and $H_B$ respectively (as usual, $U_A\otimes
U_B(\cdot)U_A^\dag\otimes U_B^\dag$ is called a local unitary
operation). Does there exist non-unitary local channel that still
transforms MES into MES? That is the main purpose of Section 2. We
show that  local channel preserves MES if and only if it is a
local unitary operation.

Note that entanglement measures reach the maximum only at MES in
general. For example, the entanglement of formation, concurrence,
distillable entanglement \cite{Bennett4} and the relative entropy
of entanglement \cite{Vedral,Miranowicz} are in such a situation
\cite{Feishaoming}. But there exists entanglement measure which is
not the case. For example, the \emph{Schmidt number}
\cite{Terhal,Sanpera,Sperling} is an entanglement measure that
reaches the maximum at MES, but not vice versa. Section 3 concerns
local channels preserving Schmidt number.

\section{Local channels preserving the maximal entanglement}

Recall that, a bipartite state is called a \emph{maximally
entangled state}(MES) if it archives the greatest  entanglement
for a certain entanglement measure (such as entanglement of
formation \cite{Wootters2,Gao}, concurrence
\cite{Wootters2,Gao,Hou7}, etc.). For an $m\otimes n$ ($m\leq n$)
system, a pure state $|\psi\rangle$ is a MES if and only if
$\rho_A=\frac{1}{m}I_a$ \cite{Horodeckip}, where $\rho_A$ is the
reduced state of $\rho=|\psi\rangle\langle\psi|$ with respect to
subsystem A. Equivalently, $|\psi\rangle$ is a MES if and only if
\begin{eqnarray}
|\psi\rangle=\frac{1}{\sqrt{m}}\sum_{i=1}^m|i\rangle|i'\rangle,\label{1}
\end{eqnarray}
where $\{|i\rangle\}$ is an orthonormal basis of $H_A$ and
$\{|i'\rangle\}$ is an orthonormal set of $H_B$. For example, the
well-known EPR states are maximally entangled pure states. MES was
discussed by several researchers (see
Refs.~\cite{Verstraete,Feishaoming,Cava} for detail). It is proved
in Ref.~\cite{Cava} that any MES in a $d\otimes d$ system is pure.
 It is worth mentioning that, very recently, Li \emph{et al.} showed
in Ref.~\cite{Feishaoming} that the maximal entanglement can also
exist in mixed states for $m\otimes n$ systems with $n\geq 2m$ (or
$m\geq 2n$). A characterization  of MES is proposed
\cite{Feishaoming}: An $m\otimes n$ ($n\geq 2m$) bipartite mixed
state $\rho$ is maximally entangled if and only if
\begin{eqnarray}
\rho=\sum\limits_{k}p_k|\psi_k\rangle\langle\psi_k|,~~\sum\limits_kp_k=1,~p_k\geq0,
\end{eqnarray}
where $|\psi_k\rangle$s are maximally entangled pure states with
\begin{eqnarray}
|\psi_k\rangle=\frac{1}{\sqrt{m}}\sum_{i=1}^m|i\rangle|i_k'\rangle,
\end{eqnarray}
$\{|i\rangle\}$ is an orthonormal basis of $H_A$ and
$\{|i_k'\rangle\}$ is an orthonormal set of $H_B$, satisfying
$\langle i_s'|j_t'\rangle=\delta_{ij}\delta_{st}$. Symmetrically,
if $m\geq 2n$, then $\rho$ is maximally entangled if and only if
\begin{eqnarray}
\rho=\sum\limits_{k}p_k|\phi_k\rangle\langle\phi_k|,~~\sum\limits_kp_k=1,~p_k\geq0,
\end{eqnarray}
where $|\phi_k\rangle$s are maximally entangled pure states with
\begin{eqnarray}
|\phi_k\rangle=\frac{1}{\sqrt{n}}\sum_{i=1}^n|i_k\rangle|i'\rangle,
\end{eqnarray}
$\{|i_k\rangle\}$ is an orthonormal set of $H_A$ satisfying
$\langle i_s|j_t\rangle=\delta_{ij}\delta_{st}$ and
$\{|i'\rangle\}$ is an orthonormal basis of $H_B$. \if For
example, in a $2\otimes 4$ system, let
$|\psi_1\rangle=\frac{1}{\sqrt{2}}(|0\rangle|0'\rangle+|1\rangle|1'\rangle)$
and
$|\psi_2\rangle=\frac{1}{\sqrt{2}}(|0\rangle|2'\rangle+|1\rangle|3'\rangle)$.
Then
\begin{eqnarray*}
\rho=p_1|\psi_1\rangle\langle\psi_1|+p_2|\psi_2\rangle\langle\psi_2|
\label{2}
\end{eqnarray*}
is a maximally entangled mixed state for any probability
distribution $\{p_i\}$.  In a $4\otimes 2$ system, let
$|\phi_1\rangle=\frac{1}{\sqrt{2}}(|0\rangle|0'\rangle+|1\rangle|1'\rangle)$
and
$|\phi_2\rangle=\frac{1}{\sqrt{2}}(|2\rangle|0'\rangle+|3\rangle|1'\rangle)$.
Then
\begin{eqnarray*}
\varrho=p_1|\phi_1\rangle\langle\phi_1|+p_2|\phi_2\rangle\langle\phi_2|
\end{eqnarray*}
is a maximally entangled mixed state for any probability
distribution $\{p_i\}$.\fi

Now we turn to our main result.\\

\noindent{\bf Theorem 2.1}\quad
A local channel preserves MES if and only if it is a local unitary operation.\\

Before proceeding to show Theorem 2.1, we recall a useful
auxiliary result on channels preserving pure states.\\

\noindent{\bf Lemma 2.2} \cite{Cubitt} \quad Let $H$ be a complex
separable Hilbert space with $\dim H\leq+\infty$, and let
$\Lambda(\cdot)=\sum\limits_{i=1}^NX_i(\cdot)X_i^\dag$ be a
channel on the quantum system described by $H$. Then $\Lambda$
transforms pure states into pure states if and only if one of the
following is true:

(1) $\Lambda$ is an isometric operation;

(2) There exists a pure state $|\omega\rangle$, such that $\Lambda(A)={\rm Tr}(A)|\omega\rangle\langle\omega|$.

\if false
 {\sl Proof.}\quad We only need to check the
`only if' part. Let $|\psi\rangle$ be a pure state. Then
\begin{eqnarray}
\Lambda(|\psi\rangle\langle\psi|)=\sum\limits_{i=1}^N
X_i|\psi\rangle\langle\psi|X_i^\dag=|\eta_\psi\rangle\langle\eta_\psi|
\end{eqnarray}
for some pure state $|\eta_\psi\rangle$. It is easy to check that
for any given pair $(k,l)$ and any pure state $|\psi\rangle$,
$X_k|\psi\rangle$ and $X_l|\psi\rangle$ are linearly dependent.

If $X_k$ and $X_l$ are linearly dependent for every pair $(k,l)$,
then  there exists an operator $X$ on $H$ and complex numbers
$\alpha_i$s, such that $ X_i=\alpha_iX,\quad i=1, 2, \dots, N$.
Let $ \check{X}=\sqrt{\sum_i|\alpha_i|^2}X,$ then
$\Lambda(\cdot)=\check{X}(\cdot)\check{X}^\dag,$ which leads to
$\check{X}^\dag\check{X}=I$. This is the case (1).

If  $X_k$ and $X_l$ are linearly independent for some pair
$(k,l)$, then we can obtain ${\rm Ker} X_k\cap{\rm Ker} X_l\neq
{\rm Ker} X_k$ and ${\rm Ker} X_k\cap{\rm Ker} X_l\neq {\rm Ker}
X_l$, where ${\rm Ker}(A)$ denotes the null space of $A$,
$A\in\mathcal{B}(H)$. That is, there exist $|x_0\rangle$,
$|y_0\rangle\in H$ such that $X_k|x_0\rangle\neq0$ and
$X_l|y_0\rangle\neq0$ with $X_k|y_0\rangle=X_l|x_0\rangle=0$. Let
$M$ be the space spanned by ${\rm Ker} X_k$ and ${\rm Ker} X_l$,
then both $X_kP_M$ and $X_lP_M$ are rank-one operators from $M$
into $H$ and ${\rm rang} X_kP_M={\rm rang} X_lP_M$, where $P_M$
stands for the projection on $M$ and ${\rm rang}(\cdot)$ denotes
the range of the operator. We assert that $M=H$. Or else, we have
${\rm rank}(X_k)>1$ or ${\rm rank}(X_l)>1$. With no loss of
generality, we assume that ${\rm rank}(X_k)>1$. Then there exists
$|x\rangle\in H\setminus{M}$ so that $X_k|x\rangle$ and $X_k
|x_0\rangle$ are linearly independent, which leads to
$X_k(|x\rangle+|x_0\rangle)$ and $X_l(|x\rangle+|x_0\rangle)$ are
linearly independent, it is impossible. Hence, $M=H$, i.e.,
 $X_k$ and $X_l$ are rank-one operators with the same
range. Consequently, $X_i$ ($i=1,\ldots N$) are  rank-one
operators with the same range. Thus the (2) holds true.
\hfill$\square$
\fi
In order to prove Theorem 2.1, the following proposition is also needed.\\

\noindent{\bf Proposition 2.3}\quad Let $H_A\otimes H_B$ be a
complex Hilbert space that describes a bipartite quantum system
A+B and let $\Lambda_{b}$ be a quantum channel on subsystem B.
Then $I_a\otimes\Lambda_b$ preserves MES if and only if
$\Lambda_b$ is a unitary operation, i.e.,
there exists a unitary operator $U$ on $H_B$ such that $\Lambda_{b}(\cdot)=U(\cdot)U^\dag$.\\

\noindent{\it Proof}\quad The `if' part is clear. It remains to
show the `only if' part. Suppose that $\dim H_A=m$, $\dim H_B=n$.
Let $\rho=|\psi\rangle\langle\psi|$ be a maximally entangled pure
state. In the following, it will be shown that $\Lambda_b$
preserves pure states and the range of $\Lambda_b$ contains linear
independent states. Then one can finish the proof by Proposition
2.2.

{\sl Case 1.}\quad $m\leq n$. Suppose that the rank of
$(I_a\otimes \Lambda_b)\rho$ is $t$.  Let
$|\psi\rangle=\sum_i\frac{1}{\sqrt{m}}|i\rangle|i'\rangle$ as in
Eq.{(\ref{1})}. Then
\begin{eqnarray}
\begin{array}{rcl}&&
\sum\limits_{i,j}|i\rangle\langle j|\otimes\Lambda_b(|i'\rangle\langle j'|)\\
&=&\sum_{i,j}|\xi_i\rangle\langle\xi_j|\otimes\sum\limits_{s=1}^tp_s|\xi_{i(s)}'\rangle\langle\xi_{j(s)}'|,
\end{array}
\end{eqnarray}
where $\{|\xi_i\rangle\}$ is an orthonormal basis of $H_A$,
$\{|\xi_{i(s)}'\rangle\}$ is an orthonormal set of $H_B$, and
where $\{p_s\}$ is a probability distribution. For an arbitrary
element $|i_0\rangle$  from $\{|i\rangle\}$, define a map
$\phi_a:\mathcal{B}(H_A)\rightarrow\mathcal{B}(H_A)$ by
\begin{eqnarray*}
\phi_a(\cdot)=|i_0\rangle\langle i_0|(\cdot)|i_0\rangle\langle
i_0|
\end{eqnarray*}
and let
\begin{eqnarray*}
|\xi_i\rangle=U_a|i\rangle,
\end{eqnarray*}
$i=1$, 2, $\dots$, $m$, $U_a=[u_{ij}]$. Then, on one hand, we have
\begin{eqnarray*}
\begin{array}{rcl}&&(\phi_a\otimes I_b)
(\sum_{i,j}|\xi_i\rangle\langle\xi_j|\otimes\sum\limits_{s=1}^tp_s|\xi_{i(s)}'\rangle\langle\xi_{j(s)}'|)\\
&=&(\phi_a\otimes I_b) (\sum\limits_{i,j}U_a|i\rangle\langle
j|U_a^\dag
\otimes\sum\limits_{s=1}^tp_s|\xi_{i(s)}'\rangle\langle\xi_{j(s)}'|)\\
&=&\sum\limits_{i,j}|i_0\rangle\langle i_0|\otimes(u_{i_0i}\bar{u}_{i_0j}\sum\limits_{s=1}^tp_s|\xi_{i(s)}'\rangle\langle\xi_{j(s)}'|)\\
&=&|i_0\rangle\langle
i_0|\otimes\sum\limits_{i,j}u_{i_0i}\bar{u}_{i_0j}
\sum\limits_{s=1}^tp_s|\xi_{i(s)}'\rangle\langle\xi_{j(s)}'|\\
&=&|i_0\rangle\langle
i_0|\otimes\sum\limits_{s=1}^tp_s|w_{i_0(s)}'\rangle\langle
w_{i_0(s)}'|,
\end{array}
\end{eqnarray*}
where
$|w_{i_0(s)}'\rangle=\sum\limits_iu_{i_0i}|\xi_{i(s)}'\rangle$. On
the other hand,
\begin{eqnarray*}
\begin{array}{rcl}&&(\phi_a\otimes I_b)
(\sum\limits_{i,j}|i\rangle\langle j|\otimes\Lambda_b(|i'\rangle\langle j'|))\\
&=&|i_0\rangle\langle i_0|\otimes\Lambda_b(|i_0'\rangle\langle
i_0'|).
\end{array}
\end{eqnarray*}
As a result
\begin{eqnarray*}
\Lambda_b(|i_0'\rangle\langle i_0'|)
=\sum\limits_{s=1}^tp_s|w_{i_0(s)}'\rangle\langle w_{i_0(s)}'|.
\end{eqnarray*}
Observing that
$\langle\xi_{i(s)}'|\xi_{j(t)}'\rangle=\delta_{ij}\delta_{st}$ and
$U_a$ is unitary, we have $\{\Lambda_b(|i'\rangle\langle i'|)\}$
is a set of mutually orthogonal rank-$t$ density operators. Let
$|v_i'\rangle$ be an orthonormal basis of $H_B$. Then the rank of
$\Lambda_b(\frac{I_b}{n})=\Lambda_b(\frac{\sum_i|v_i'\rangle\langle
v_i'|}{n})$ is $tn$, which implies $t=1$ as desired. Thus
$\Lambda_b$ preserves pure states.

{\sl Case 2.}\quad $n\leq m$. Note that a state $\varrho$ in an
$m\otimes n$ system is maximally entangled if and only if
\begin{eqnarray*}
\varrho=\sum\limits_{k}p_k|\phi_k\rangle\langle\phi_k|,~~\sum\limits_kp_k=1,~p_k\geq0,
\end{eqnarray*}
where $|\phi_k\rangle$s are maximally entangled pure states with
\begin{eqnarray*}
|\phi_k\rangle=\frac{1}{\sqrt{n}}\sum_{i=1}^n|i_k\rangle|i'\rangle,
\end{eqnarray*}
$\{|i_k\rangle\}$ is an orthonormal set of $H_A$ satisfying
$\langle i_s|j_t\rangle=\delta_{ij}\delta_{st}$ and
$\{|i'\rangle\}$ is an orthonormal basis of $H_B$. Let
$|\psi\rangle=\sum_i\frac{1}{\sqrt{n}}|i\rangle|i'\rangle$ be a
maximally entangled pure state, $\{|i\rangle\}_{i=1}^{n}$ is a
orthonormal set of $H_A$. Let $\{|i\rangle\}_{i=1}^{m}$ be an
orthonormal  basis of $H_A$ extended from
$\{|i\rangle\}_{i=1}^{n}$. Write $\rho=|\psi\rangle\langle\psi|$.
We assume that the rank of $(I_a\otimes\Lambda_b)\rho$ is $r$. We
check that $r=1$. Let
\begin{eqnarray*}
\begin{array}{rcl}&&
\sum\limits_{i,j}|i\rangle\langle j|\otimes\Lambda_b(|i'\rangle\langle j'|)\\
&=&\sum_{i,j}\sum\limits_{s=1}^rq_s|\xi_{i(s)}\rangle\langle\xi_{j(s)}|\otimes|\xi_i'\rangle\langle\xi_j'|,
\end{array}
\end{eqnarray*}
where $\{|\xi_{i(s)}\rangle\}$ is an orthonormal set of $H_A$,
$\{|\xi_{i}'\rangle\}$ is an orthonormal basis of $H_B$, and where
$\{q_s\}$ is a probability distribution. Let
\begin{eqnarray*}
|\xi_{i(s)}\rangle=U_{a(s)}|i\rangle, \quad 1\leq i\leq n
\end{eqnarray*}
\begin{eqnarray*}
U_{a(s)}|i\rangle=0, \quad i>n
\end{eqnarray*}
for some operators $U_{a(s)}$s, $s=1$, 2, $\dots$, $r$. It is
straightforward that  $U_{a(s)}^\dag U_{a(t)}=0$ when $s\neq t$.
Write $U_{a(s)}=[u_{ij}^{(s)}]$ with respect to the basis extended
from $\{|i\rangle\}$. Then
\begin{eqnarray*}
\begin{array}{rcl}&&(\phi_a\otimes I_b)
(\sum_{i,j}\sum\limits_{s=1}^rq_s|\xi_{i(s)}\rangle\langle\xi_{j(s)}|\otimes|\xi_i'\rangle\langle\xi_j'|)\\
&=&(\phi_a\otimes I_b)
(\sum\limits_{i,j}\sum\limits_{s=1}^rq_sU_{a(s)}|i\rangle\langle
j|U_{a(s)}^\dag
\otimes|\xi_{i}'\rangle\langle\xi_{j}'|)\\
&=&\sum\limits_{i,j}|i_0\rangle\langle i_0|\otimes\sum\limits_{s=1}^rq_su_{i_0i}^{(s)}\bar{u}_{i_0j}^{(s)}|\xi_{i}'\rangle\langle\xi_{j}'|\\
&=&|i_0\rangle\langle
i_0|\otimes\sum\limits_{s=1}^rq_s\sum\limits_{i,j}u_{i_0i}^{(s)}\bar{u}_{i_0j}^{(s)}
|\xi_{i}'\rangle\langle\xi_{j}'|\\
&=&|i_0\rangle\langle
i_0|\otimes\sum\limits_{s=1}^rq_s|\omega_{i_0(s)}'\rangle\langle
\omega_{i_0(s)}'|,
\end{array}
\end{eqnarray*}
where
$|\omega_{i_0(s)}'\rangle=\sum\limits_ku_{i_0k}^{(s)}|\xi_{k}'\rangle$.
It turns out that
\begin{eqnarray}
\Lambda_b(|i_0'\rangle\langle i_0'|)
=\sum\limits_{s=1}^rq_s|\omega_{i_0(s)}'\rangle\langle
\omega_{i_0(s)}'|.
\end{eqnarray}
Consequently, $\{\Lambda_b(|i'\rangle\langle i'|)\}$ is a set of
mutually orthogonal rank-$r$ density operators which indicated
that the rank of
$\Lambda_b(\frac{I_b}{n})=\Lambda_b(\frac{\sum_i|v_i'\rangle\langle
v_i'|}{n})$ is $rn$, and $r=1$. That is $\Lambda_b$ preserves pure
state. \hfill$\qed$

Similarly, one can show that $\Lambda_a\otimes I_b$ preserves MES if and only if there exists a unitary operator
$U$ acting on $H_A$ such that  $\Lambda_{a}(\cdot)=U(\cdot)U^\dag$.\\

\noindent{\bf Proof of Theorem 2.1}\quad  We only need to show the
`only if' part.

Let $\Lambda_a\otimes\Lambda_b$ be a local channel. Observe that
$\Lambda_a\otimes\Lambda_b$ can be viewed as a local operation and
classical communication (LOCC) \cite{Horodecki1} and
\begin{eqnarray*}
\begin{array}{rcl}
(\Lambda_a\otimes\Lambda_b)\rho&=&(\Lambda_a\otimes I_b)(I_a\otimes\Lambda_b)\rho\\
&=&(I_a\otimes \Lambda_b)(\Lambda_a\otimes I_b)\rho.
\end{array}
\end{eqnarray*}
Write $\rho'=(I_a\otimes\Lambda_b)\rho$ and
$\rho''=(\Lambda_a\otimes\Lambda_b)\rho$. Then $ E_f(\rho'')\leq
E_f(\rho')\leq E_f(\rho)$ for any state $\rho$ acting on
$H_A\otimes H_B$ since entanglement measure is monotonic under
LOCC \cite{Vidal,Fan}, where $E_f$ denotes the entanglement of
formation. If $\rho$ is a MES, by the assumption, one has
$E_f(\rho'')=E_f(\rho),$ and thus
\begin{eqnarray}
E_f(\rho'')=E_f(\rho')=E_f(\rho).
\end{eqnarray}
It follows that that both $\rho'$ and $\rho''$ are MES since
$E_f(\rho)$ reaches the maximum if and only if $\rho$ is a MES
\cite{Feishaoming}. Thus, by Proposition 2.3, $\Lambda_b$ is a
unitary operation. Similarly, $\Lambda_a$ is a unitary operation
as well. The proof is completed.
\hfill$\qed$

From Theorem 2.1, if a local channel $\Lambda$ preserves MES, then
it preserves entanglement measure since entanglement measure is
invariant under local unitary operation \cite{Vidal,Vidal3,Vedral}
(Here, we call $\Lambda$ preserves entanglement measure $E$ if
$E(\Lambda(\rho))=E(\rho)$).
The following is straightforward.\\

\noindent{\bf Proposition 2.3}\quad Let $E$ be an entanglement
measure that reaches maximum only at MES.
Then a local channel preserves the entanglement measure quantified by $E$ if and only if it is a local unitary operation.\\

\section{Local channels preserving the Schmidt number}
Though entanglement measures reach the maximum only at MES in
general, there exists entanglement measure which is not the case.
The \emph{Schmidt number} \cite{Terhal,Sanpera,Sperling} is one of
important entanglement measure that reaches the maximum at MES,
but not vice versa. Among  various entanglement measures, Schmidt
number deduced from the Schmidt rank~\cite{Terhal} is a universal
entanglement measure~\cite{Sperling}. This section is devoted to
local channels preserving Schmidt number.

Recall that, for the finite-dimensional case, the Schmidt number,
denoted by $r_S(\rho)$, is defined by
\cite{Terhal,Sanpera,Sperling}
\begin{eqnarray}
r_S(\rho):=\inf\{\max\limits_{i}[r_S(|\psi_i\rangle)]\},\label{3}
\end{eqnarray}
where the infimum is taken over all pure state decompositions
$\rho=\sum\limits_i p_i|\psi_i\rangle\langle\psi_i|$,
$\sum\limits_ip_i=1$, $p_i>0$, and $r_S(|\psi_i\rangle)$ is the
Schmidt rank of $|\psi_i\rangle$. If $|\psi\rangle$ is a pure
state with Schmidt decomposition $
|\psi\rangle=\sum\limits_{k=1}^{r_S(|\psi\rangle)}\lambda_k|k\rangle|k^{\prime}\rangle
$, then $r_S(|\psi\rangle)$ is called the Schmidt rank of
$|\psi\rangle$ and $\lambda_k$s are the Schmidt coefficients of
$|\psi\rangle$. The Schmidt coefficients play a minor role
compared with the Schmidt number in the quantification of
entanglement~\cite{Sperling}. Using the same scenario, one can
extend the Schmidt number to infinite-dimensional bipartite
systems. Suppose that $\dim H_A\otimes H_B=+\infty$. We define
\begin{eqnarray}
r_S(\rho):=\inf\{\sup\limits_{i}[r_S(|\psi_i\rangle)]\}.\label{4}
\end{eqnarray}
$r_S$ may be $+\infty$ whenever $\dim H_A=+\infty$ and $\dim
H_B=+\infty$.

By definitions in~Eqs.~(\ref{3})-(\ref{4}), the following
properties are straightforward: (i) A pure state is separable if
and only if its Schmidt number is 1.  (ii) If
$\rho=|\psi\rangle\langle\psi|$, $\dim H_A=m\leq\dim H_B$,
$m<+\infty$, then $r_S(|\psi\rangle)=m_0\leq m$ if and only if the
rank of the reduced state  $\text{tr}_A(\rho)$ is $m_0$. (iii)
$|\psi\rangle$ is a maximally entangled state implies
$r_S(|\psi\rangle)$ reaches the greatest value, but not vice
versa.

As usual, if $\Lambda$ satisfies $r_S(\Lambda(\rho))=r_S(\rho)$
for any state $\rho$, we call $\Lambda$ preserves the Schmidt
number. Let $1\leq r\leq\min\{\dim H_A,\dim H_B\}$ be an
arbitrarily given integer. If $\Lambda(|\psi\rangle\langle\psi|)$
is pure and $r_S(\Lambda(|\psi\rangle\langle\psi|))=r$  when
$r_S(|\psi\rangle\langle\psi|)=r$, we call $\Lambda$ preserves
pure entangled states with Schmidt number $r$. We concern with the
local channel preserving pure entangled states with Schmidt number
$r$, for a given number $r$. To our surprise, such channel must be
local isometric operation. This implies that if a local channel
preserves pure entangled states with an arbitrarily fixed Schmidt
number, then it preserves all Schmidt numbers for both pure and
mixed states.

The following are our main results of this section.\\

\noindent{\bf Theorem 3.1}\quad  Suppose that $\dim H_A\otimes
H_B\leq+\infty$ and $2\leq r\leq\min\{\dim H_A,\dim H_B\}$. Then a
local channel preserves pure states with Schmidt number $r$  if
and only if it is a local isometric operation.\\

\noindent{\bf Theorem 3.2}\quad A local channel preserves
separable pure states  if and only if it is a local  isometric
operation or it transforms any state into  pure separable states.

Just as in Section 2, before giving the proof of Theorem 3.1 and
3.2, we first treat the local channel
$I_a\otimes\Lambda_b$.\\

\noindent{\bf Proposition 3.3}\quad Let $\Lambda_{b}$ be a quantum
channel on subsystem B, and $r$ be a fixed positive integer no
larger than $\min\{\dim H_A,\dim H_B\}$. Then
$I_a\otimes\Lambda_b$ on the bipartite system A+B preserves pure
states with Schmidt number $r$ if and only if either (i)
$\Lambda_b$ is an isometric operation or (ii) $\Lambda_b$
transforms any states into pure states. In case (ii),
$I_a\otimes\Lambda_b$ sends any states to pure separable states.
Consequently, the (ii) can't occur when $r\neq 1$.

{\bf Proof.}\quad  For simplicity, we suppose that $\dim H_A=m$,
$\dim H_B=n$, $m\leq n\leq+\infty$. Let
$\rho=|\psi\rangle\langle\psi|$ with
$|\psi\rangle=\sum\limits_{i=1}^{r}\lambda_i|i\rangle|i'\rangle$,
$\lambda_i>0$, $1\leq i\leq r$.

The `if' part.  Now let $\Lambda_b(\cdot)=X(\cdot)X^\dag$ with $X$
is an isometric operator. Write $X|i'\rangle=|\eta_i'\rangle$.
Then $I_a\otimes
X|\psi\rangle=\sum\limits_{i=1}^{r}\lambda_i|i\rangle\otimes
X|i'\rangle
=\sum\limits_{i=1}^{r}\lambda_i|i\rangle\otimes|\eta_i'\rangle$.
Write $|\phi\rangle=I_a\otimes X|\psi\rangle$. It is easy to check
that ${\rm
Tr}_A(|\phi\rangle\langle\phi|)=\sum\limits_i\lambda_i^2|\eta_i'\rangle\langle\eta_i'|$
is of $r$-rank, which reveals that
$r_S((I_a\otimes\Lambda_b)(|\psi\rangle\langle\psi|))=r$.

If the case (ii) occurs, let $\Lambda_b(\cdot)=\sum_{i=1}^{N}
X_i(\cdot)X_i^\dag$. By Proposition 2.2, $X_k=a_k|x\rangle\langle
y_k|$ for some $|x\rangle$, $|y_k\rangle\in H_B$. It follows that,
for $\rho=|\psi\rangle\langle\psi|$ with
$|\sigma\rangle=\sum\limits_{i=1}^{r_{s}(|\psi\rangle)}\lambda_i|i\rangle|i'\rangle$,
$\lambda_i>0$,
\begin{eqnarray*}
\begin{array}{rcl}
&&(I_a\otimes\Lambda_b)\sigma\\
&=&\sum\limits_{i,j}\lambda_i\lambda_j|i\rangle\langle j|\otimes\sum\limits_k|a_k|^2\alpha_{ik}\overline{\alpha_{jk}}|x\rangle\langle x|\\
&=&\sum\limits_k|w_k\rangle\langle w_k|\otimes|x\rangle\langle x|,
\end{array}
\end{eqnarray*}
where $a_{ik}=\langle y_k|i'\rangle$,
$|w_k\rangle=\sum\limits_i\lambda_ia_k\alpha_{ik}|i\rangle$ (Here
$|w_k\rangle$ may not be  normalized).  It is easy to see that
$\sum\limits_k|w_k\rangle\langle w_k|\otimes|x\rangle\langle x|$
is separable and its Schmidt number is $1$, i.e.,
$I_a\otimes\Lambda_b$ sends all states into pure stases.  So this
case cannot occur if $r\neq 1$.

The `only if' part.  Observe that
\begin{eqnarray*}
\begin{array}{rcl}&&
\sum\limits_{i,j}\lambda_i\lambda_j|i\rangle\langle j|\otimes\Lambda_b(|i'\rangle\langle j'|)\\
&=&\sum_{i,j}\delta_i\delta_j|\xi_i\rangle\langle\xi_j|\otimes|\xi_{i}'\rangle\langle\xi_{j}'|,
\end{array}
\end{eqnarray*}
where $\{|\xi_i\rangle\}_{i=1}^{r}$ and
$\{|\xi_{i}'\rangle\}_{i=1}^{r}$ are  orthonormal sets of $H_A$
and $H_B$, respectively. $\sum\limits_{i=1}^{r}\delta_i^2=1$. Let
$\{|i\rangle\}_{i=1}^m$ be a basis of $H_A$ extended from
$\{|i\rangle\}_{i=1}^{r}$.   Define partial isometry operator
$U_a$ on $H_A$ by
\begin{eqnarray*}
U_a|i\rangle=|\xi_i\rangle, \quad 1\leq i\leq r
\end{eqnarray*}
\begin{eqnarray*}
U_a|i\rangle=0, \quad r<i\leq m.
\end{eqnarray*}
Write $U_a=[u_{ij}]$, where $u_{ij}=\langle i|U_a|j\rangle$. For
every $|i\rangle$ from $\{|i\rangle\}_{i=1}^m$, define
$\phi_a:\mathcal{T}(H_A)\rightarrow\mathcal{T}(H_A)$ be a map
defined by
\begin{eqnarray*}
\phi_a(\cdot)=|i_0\rangle\langle i_0|(\cdot)|i_0\rangle\langle
i_0|.
\end{eqnarray*}
Consequently, on one hand, we have
\begin{eqnarray*}
\begin{array}{rcl}&&(\phi_a\otimes I_b)
(\sum_{i,j}\delta_i\delta_j|\xi_i\rangle\langle\xi_j|\otimes|\xi_{i}'\rangle\langle\xi_{j}'|)\\
&=&(\phi_a\otimes I_b)
(\sum\limits_{i,j}\delta_i\delta_jU_a|i\rangle\langle j|U_a^\dag
\otimes|\xi_{i}'\rangle\langle\xi_{j}'|)\\
&=&\sum\limits_{i,j}\delta_i\delta_j|i_0\rangle\langle i_0|\otimes(u_{i_0i}\overline{u_{i_0j}}|\xi_{i}'\rangle\langle\xi_{j}'|)\\
&=&|i_0\rangle\langle
i_0|\otimes\sum\limits_{i,j}\delta_i\delta_ju_{i_0i}\overline{u_{i_0j}}
|\xi_{i}'\rangle\langle\xi_{j}'|\\
&=&|i_0\rangle\langle i_0|\otimes|w_{i_0}'\rangle\langle
w_{i_0}'|,
\end{array}
\end{eqnarray*}
where
$|w_{i_0}'\rangle=\sum\limits_i\delta_iu_{i_0i}|\xi_{i}'\rangle$
($|w_{i_0}'\rangle$ may not be normalized). On the other hand,
\begin{eqnarray*}
\begin{array}{rcl}&&(\phi_a\otimes I_b)
(\sum\limits_{i,j}\lambda_i\lambda_j|i\rangle\langle j|\otimes\Lambda_b(|i'\rangle\langle j'|))\\
&=&\lambda_{i_0}^2|i_0\rangle\langle
i_0|\otimes\Lambda_b(|i_0'\rangle\langle i_0'|).
\end{array}
\end{eqnarray*}
As a result
\begin{eqnarray}
\lambda_{i_0}^2\Lambda_b(|i_0'\rangle\langle i_0'|)
=|w_{i_0}'\rangle\langle w_{i_0}'|.
\end{eqnarray}
Using Proposition 2.2, one can finish the proof. \hfill$\square$

Similarly, one can show that $\Lambda_a\otimes I_b$ preserves an
arbitrarily fixed Schmidt number for pure entangled states if and
only if either  $\Lambda_a$ is an isometric operation or
$\Lambda_a$ transforms any states into pure states.

{\bf Proof of Theorem 3.1}\quad  We only need to show the `only
if' part.

Let $\Lambda_a\otimes\Lambda_b$ be a local channel on the
bipartite system A+B. Observe that $\Lambda_a\otimes\Lambda_b$ can
be viewed as a local operation and classical communication (LOCC).
Write $\rho'=(I_a\otimes\Lambda_b)\rho$ and
$\rho''=(\Lambda_a\otimes\Lambda_b)\rho$. Then $ r_S(\rho'')\leq
r_S(\rho')\leq r_S(\rho)$ for any state $\rho$ acting on
$H_A\otimes H_B$ since $r_S$ is monotonic decreasing under
LOCC~\cite{Terhal}. If $r_S(\rho)=r\geq2$ ($r$ may be $+\infty$),
by the assumption, one has $r_S(\rho'')=r_S(\rho)=r,$ and thus
\begin{eqnarray}
r_S(\rho'')=r_S(\rho')=r_S(\rho).
\end{eqnarray}
It follows from Proposition 3.3 that $\Lambda_b$ is an isometric
operation. Similarly, $\Lambda_a$ is an isometric operation as
well. The proof is completed. \hfill$\square$

At last, the Theorem 3.2 can be followed directly from Proposition
3.3 .

\end{document}